\documentclass[aps,prl,twocolumn,showpacs,superscriptaddress]{revtex4-1}

\usepackage[english]{babel}
\usepackage{color}
\usepackage{amsmath, amsfonts, amssymb}
\usepackage{graphicx}

\usepackage{color}
\definecolor{Blue}{rgb}{0.00, 0.00, 1.00}
\definecolor{Red}{rgb}{1.00, 0.00, 0.00}

\newcommand{\be}{\begin{equation}}
\newcommand{\ee}{\end{equation}}
\newcommand{\bea}{\begin{eqnarray}}
\newcommand{\eea}{\end{eqnarray}}

\newcommand{\rmd}{\mathrm{d}}

\begin{document}

\title{Quantitative scaling of magnetic avalanches}
\author{G. Durin}
\affiliation{Istituto Nazionale di Ricerca Metrologica, Strada delle Cacce 91, 
10135 Torino, Italy}
\affiliation{ISI Foundation, Via Alassio 11/c, 10126 Torino, Italy}
\author{F. Bohn}
\author{M.~A. Corr\^{e}a}
\affiliation{Departamento de F\'{\i}sica Te\'{o}rica e Experimental, 
Universidade Federal do Rio Grande do Norte, 59078-900 Natal, RN, Brazil}
\author{R.~L. Sommer}
\affiliation{Centro Brasileiro de Pesquisas F\'{i}sicas, Rua Dr.\ Xavier Sigaud 150, Urca, 
22290-180 Rio de Janeiro, RJ, Brazil}
\author{P. Le Doussal}
\affiliation{CNRS-Laboratoire de Physique Th\'eorique de l'Ecole Normale 
Sup\'erieure, 24 rue Lhomond, 75005 Paris, France}
\author{K.~J. Wiese}
\affiliation{CNRS-Laboratoire de Physique Th\'eorique de l'Ecole Normale 
Sup\'erieure, 24 rue Lhomond, 75005 Paris, France}

\begin{abstract}
We provide the first quantitative comparison between Barkhausen 
noise experiments and recent predictions from the theory of avalanches 
for pinned interfaces, both in and beyond mean-field. 
We study different classes of soft magnetic materials: polycrystals and 
amorphous samples, characterized by long-range and short-range elasticity, 
respectively; both for thick and thin samples, {\it i.e.} with and without 
eddy currents. 
The temporal avalanche shape at fixed size, and observables related to the joint 
distribution of sizes and durations are analyzed in detail. Both long-range 
and short-range samples with no eddy currents are fitted extremely well 
by the theoretical predictions. In particular, the short-range samples 
provide 
the first reliable test of the theory beyond mean-field. The thick samples show 
systematic deviations from the scaling theory, providing unambiguous 
signatures for the presence of eddy currents.
\end{abstract}

\pacs{89.75.Da, 75.60.Ej, 75.60.Ch, 75.70.Ak} 

\keywords{Magnetic systems, Magnetization dynamics, Barkhausen noise, Ferromagnetic films}

\maketitle

Barkhausen noise in soft magnets originates from the jerky motion of 
magnetic domain walls (DWs), and is characterized by scale-free power-law 
distributions of magnetization 
jumps \cite{BAR-19,
UrbachMadisonMarkert1995,KimChoeShin2003,
RepainBauerJametFerreMouginChappertBernas2004,DUR-06,
PapanikolaouBohnSommerDurinZapperiSethna2011}. 
It is the earliest and most scrutinized probe for avalanche motion, an 
ubiquitous phenomenon present in other systems such as fluid contact-line 
depinning 
\cite{LeDoussalWieseMoulinetRolley2009,MoulinetGuthmannRolley2002},
brittle fracture fronts 
\cite{BonamySantucciPonson2008,LaursonIllaSantucciTallakstadyAlava2013}, and 
pinned vortex lines {\cite{SHA-15}}. In all these systems the motion of 
an overdamped elastic interface (of internal dimension $d$) driven in a 
quenched medium was proposed as an efficient mesoscopic description. 
However, until now, analytical predictions, 
allowing  for a detailed comparison with experiments, 
have been scarce, due to the difficulty in treating {\it collective 
discontinuous} jumps in  presence of many metastable states.

Toy models have thus been developed, capturing essential features 
at the level of mean-field (MF). One celebrated example is the ABBM model, 
where the domain wall is modeled as a single point in an ``effective'' random 
force landscape performing a (biased) Brownian motion
\cite{AlessandroBeatriceBertottiMontorsi1990,
AlessandroBeatriceBertottiMontorsi1990b,Colaiori2008}.
Refinements based on infinite-range models were later 
proposed \cite{DSFisher1998}, leading to similar physics 
\cite{ZapperiCizeauDurinStanley1998,Colaiori2008}.
These MF toy models predict an avalanche-size distribution $P(S) \sim 
S^{-\tau}$ with $\tau=\tau_{\rm MF}=3/2$ and a duration distribution $P(T) \sim 
T^{-\alpha}$ 
with $\alpha=\alpha_{\rm MF}=2$. 

The theory of interface depinning provides a predictive universal   
framework for the avalanche statistics. It involves two 
independent exponents, the roughness exponent $\zeta$ and the dynamical 
exponent $z$. The distribution 
exponents $\alpha$ and $\tau$ were conjectured from scaling, as in the 
Narayan-Fisher (NF) conjecture $\tau=2-\frac{\mu}{d+\zeta}$ and $\alpha = 
1+\frac{d+\zeta-\mu}{z}$, 
where $\mu$ describes the range of interactions.
The upper-critical dimension at which $\zeta=0$ and below which  
mean-field models fail is $d_{\rm uc}=4$ for short-range elasticity (SR, $\mu=2$) 
and $d_{\rm uc}=2$ for long-range elasticity (LR, $\mu=1$)
\cite{ZapperiCizeauDurinStanley1998}.

In Barkhausen noise experiments, two distinct families of samples were 
identified, consistent with these  predictions 
\cite{DurinZapperi2000}.
In polycrystaline materials, the DW experiences strong anisotropic crystal 
fields leading to LR elasticity. In the $D=d+1=2+1$ geometry 
this system behaves according to MF theory. 
In amorphous samples, SR elasticity prevails over a negligible LR elasticity, 
and the avalanche exponents  agree with the NF prediction. 
In both cases, a relevant role is played by the demagnetizing field, 
which acts as a cutoff for large  avalanches \cite{DurinZapperi2000}. 
These mean-field predictions were tested in soft magnetic thin films, where the 
retarding effects of eddy currents (ECs) of bulk samples are 
negligible \cite{PapanikolaouBohnSommerDurinZapperiSethna2011}.

Recently, it became possible to compare theory and experiments of 
avalanches well beyond
scaling and the value of exponents. On the theoretical side, the functional 
renormalization group  of depinning was extended to calculate 
a host of avalanche observables \cite{LeDoussalWiese2008c,RossoLeDoussalWiese2009a,LeDoussalWiese2011a,
DobrinevskiLeDoussalWiese2011b,LeDoussalWiese2012a,
DobrinevskiLeDoussalWiese2014a}.
Examples are the avalanche shape at fixed size
and duration, the joint size distribution, both in mean-field and beyond,
even including  retardation effects due to eddy currents \cite{DobrinevskiLeDoussalWiese2013}. On the experimental
side, the avalanche shape was studied in magnetic systems 
\cite{PapanikolaouBohnSommerDurinZapperiSethna2011,ZAP-05}, 
 in fracture and imbibition \cite{LAU-13}. Despite these experiments,
 most of the recent predictions of the theory have not
 yet been tested quantitatively, especially not to high accuracy.

The aim of this Letter is to provide new and sensitive tests of 
these theoretical predictions in soft ferromagnets. This is possible since in 
our Barkhausen experiment we detected a high number of avalanches, getting a 
robust statistics. Diverse magnetic samples have been explored, corresponding to 
the two universality classes (LR and SR), with and without EC effects. We focus 
our attention on the avalanche shape at fixed size, and  observables linked to 
the joint distribution of sizes and 
durations. The SR and LR samples with no ECs fit extremely well with the  
theoretical predictions. In particular, the SR 
samples  for the first time provide a significant test of the theory beyond 
mean-field.
The effect of eddy currents on the scaling properties is also 
investigated.

\begin{figure}[tbt]
\includegraphics[width=8.5cm]{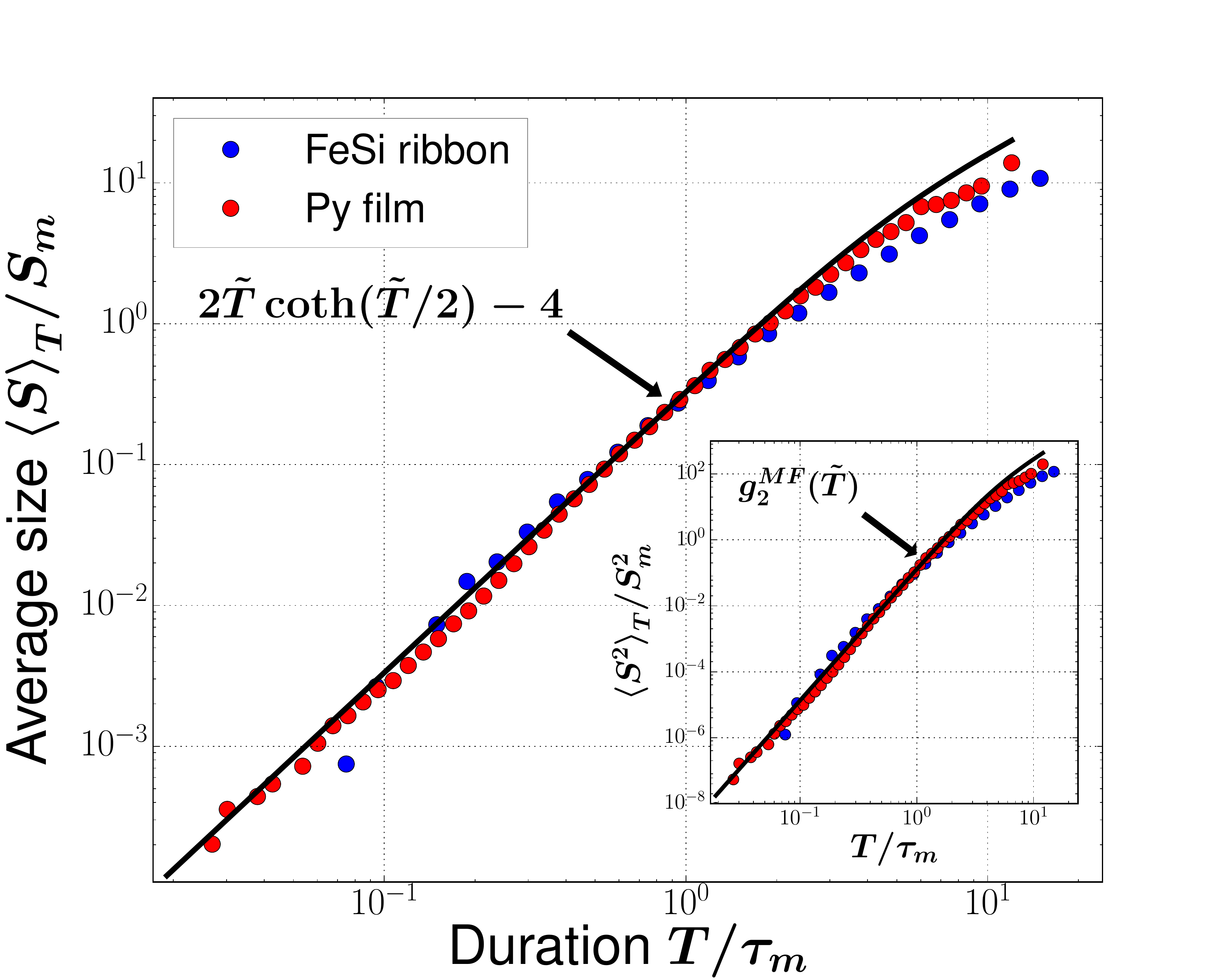}
\caption{Normalized average size $\langle S \rangle_T/S_m $ of Barkhausen 
avalanches in the FeSi ribbon (blue dots) and the Py thin film (red dots) as a 
function of the normalized duration $\tilde T = T/\tau_m$. The continuous line 
is the theoretical prediction $g_1^{\rm MF}$ of Eq.~(\ref{eq:f1_MF}). For the 
ribbon, the deviation at large durations is more 
evident due to effect of the eddy currents. The inset shows 
the second moment $\langle S^2 \rangle_T/S_m^2 $
compared to the prediction $g_2^{\rm MF}$ of Eq.~(\ref{eq:f1_MF}).}
\label{fig:aveS_MF}
\end{figure}%

We start by presenting the predictions from the theory of elastic interfaces 
that we aim to test
\cite{LeDoussalWiese2011a,LeDoussalWiese2012a,DobrinevskiLeDoussalWiese2011b,DobrinevskiLeDoussalWiese2014a}. 
These predictions are calculated for avalanches following an infinitesimal 
increase in the field (kick). They also apply to the stationary, quasi-static 
regime in the limit of slow driving, as performed in experiments 
\cite{LeDoussalWiese2012a}.
The interface model involves a (small) mass $m^2$, which flattens the interface 
beyond the scale $1/m$, playing the same role as the demagnetizing field in 
setting the cutoff scale. 
Associated to the two independent exponents $\zeta$ and $z$ are two independent 
scales $S_m \simeq m^{-d + \zeta}$ and $\tau_m \simeq m^{-z}$, for sizes and 
durations. 
The size scale $S_m$ can be directly measured in the experiments as
\bea
S_m = \frac{\langle S^2 \rangle}{2 \langle S \rangle} ,
\label{eq:Sm}
\eea
where $\langle...\rangle$ denotes expectation values w.r.t.\ $P(S)$. On 
the other hand, the time scale $\tau_m$ cannot be determined analytically, and 
has to be guessed from data, as we explain later. We denote by $u(x,t)$ the 
displacement field of the interface, $x \in \mathbb{R}^d$ and by $\dot {\sf 
u}(t) = \int 
d^dx \; \dot u(x,t)$ the time derivative of the total swept area.  
The avalanche size is $S=\int_0^T dt \; \dot {\sf u}(t)$. 

The simplest observables we can consider are the moments of the average size at 
fixed duration
\bea
\langle S^n \rangle_T = (S_m)^n g_n(T/\tau_m), \quad g_n(\tilde T) 
\simeq_{\tilde T \to 0} c_n \tilde T^{n \gamma},
\label{eq:aveS}
\eea  
whose universal behavior for small avalanches defines the exponent $\gamma$. 
Here, $\tilde T = T/\tau_m$ is the rescaled avalanche duration. In mean field, one 
finds
\bea \label{eq:gamma_MF}
\gamma^{\rm MF}=2, \quad c^{\rm MF}_1=\frac{1}{3}, \quad 
c_2^{\rm MF}=\frac{2}{15},
\eea 
and the scaling functions are
\bea
\label{eq:f1_MF}
 g_1^{\rm MF}(\tilde T)&=&2 \tilde T \coth(\tilde T/2)-4, \\
 g_2^{\rm MF}(\tilde T)&=& 2 \tilde T \text{csch}^2\!\left(\frac{\tilde T}{2}\right) \Big[\tilde T (\cosh(\tilde T)+2)-3 
\sinh (\tilde T)\Big]. \nonumber
\eea 
Beyond mean-field, one finds
\bea
\gamma = \frac{d+\zeta}{z}, \quad c_1 = c_1^{\rm MF} + \frac{11-3 \gamma_{\rm E} - \ln 
4}{81} ~\epsilon,
\label{eq:gamma}
\eea 
where the expression for $c_1$ has been calculated to first order in $\epsilon 
= d_{\rm uc} - d$.
This leads to $c_1 \approx 0.528$ for $\epsilon=2$, as in the case of $\mu=2$ 
and $d=2$ ($\gamma_{\rm E} \approx 0.577$).

\begin{figure}[tb]
\includegraphics[width=8.5cm]{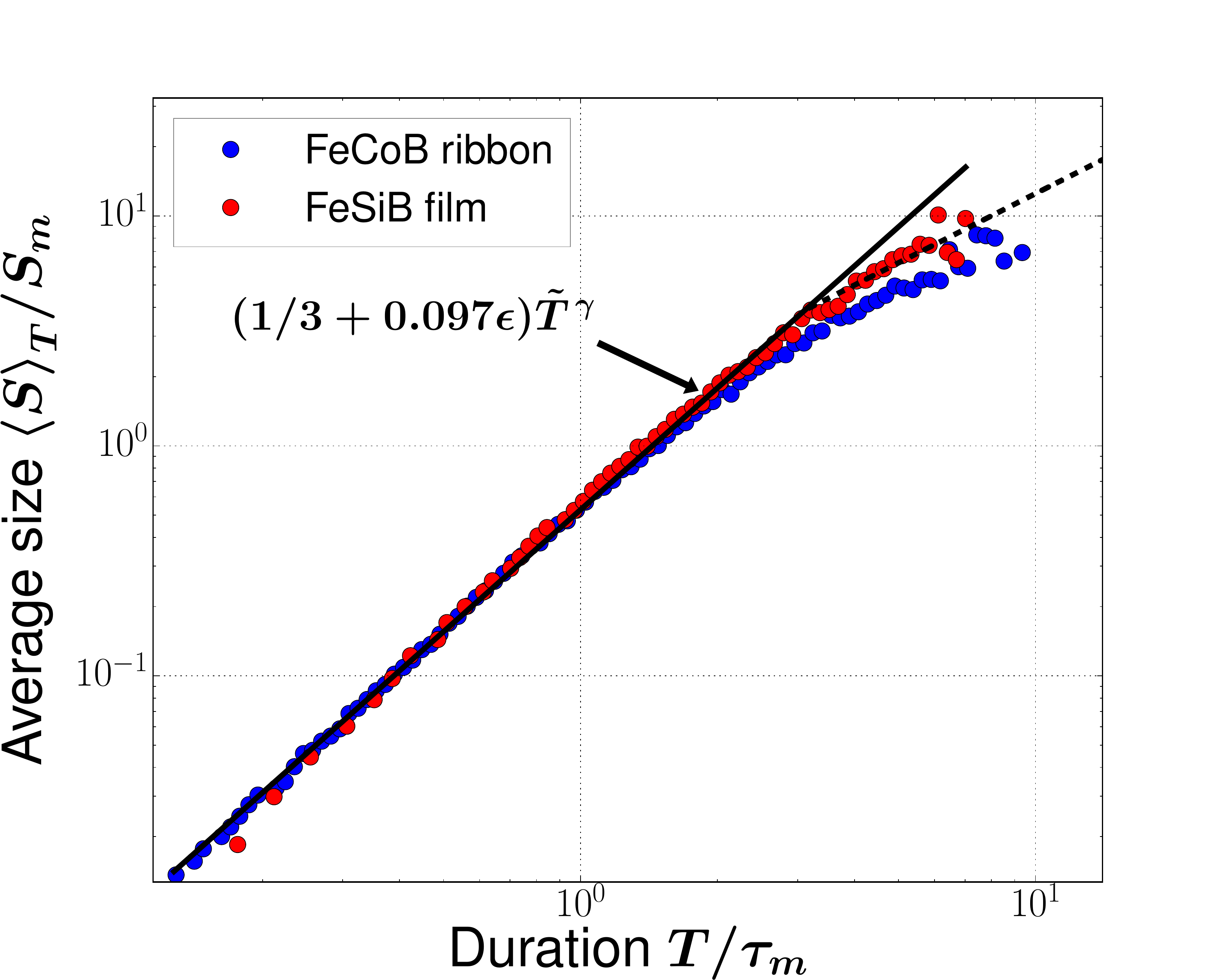}
\caption{Normalized average size $\langle S \rangle_T/S_m $ of 
avalanches in the FeCoB ribbon (blue dots) and the FeSiB thin film (red dots) as 
a function of $T/\tau_m$. The continuous line is the theoretical prediction of 
Eq.~(\ref{eq:gamma}), with $\epsilon=2$, so that $\langle S \rangle_T /S_m \sim 
0.528 \ (T/\tau_m)^{\gamma}$, with $\gamma \sim 1.76$. The ribbon shows a larger 
deviation due to eddy currents. A comparison with the expected linear behavior at large $\tilde T$
is indicated by the dashed line. Remarkably, this deviation occurs at sizes larger than the size cutoff $4 S_m$, {\it i.e.} for
$\langle S \rangle_T /S_m > 4.$}
\label{fig:aveS}
\end{figure}

Another observable of interest is the averaged avalanche duration at fixed size. In mean field, it
is given by 
\bea \label{eq:TvsSMF} 
\langle T \rangle_S / \tau_m = \sqrt{\pi S/S_m} ,
\eea 
which is consistent with the general expected scaling $\langle \tilde T \rangle_S \sim S^{1/\gamma}$,
with $\gamma=\gamma^{\rm MF}$. Remarkably, within mean-field, Eq.~(\ref{eq:TvsSMF}) holds 
for any value of the ratio $S/S_m$.

\begin{figure}[h]
\includegraphics[width=8.5cm]{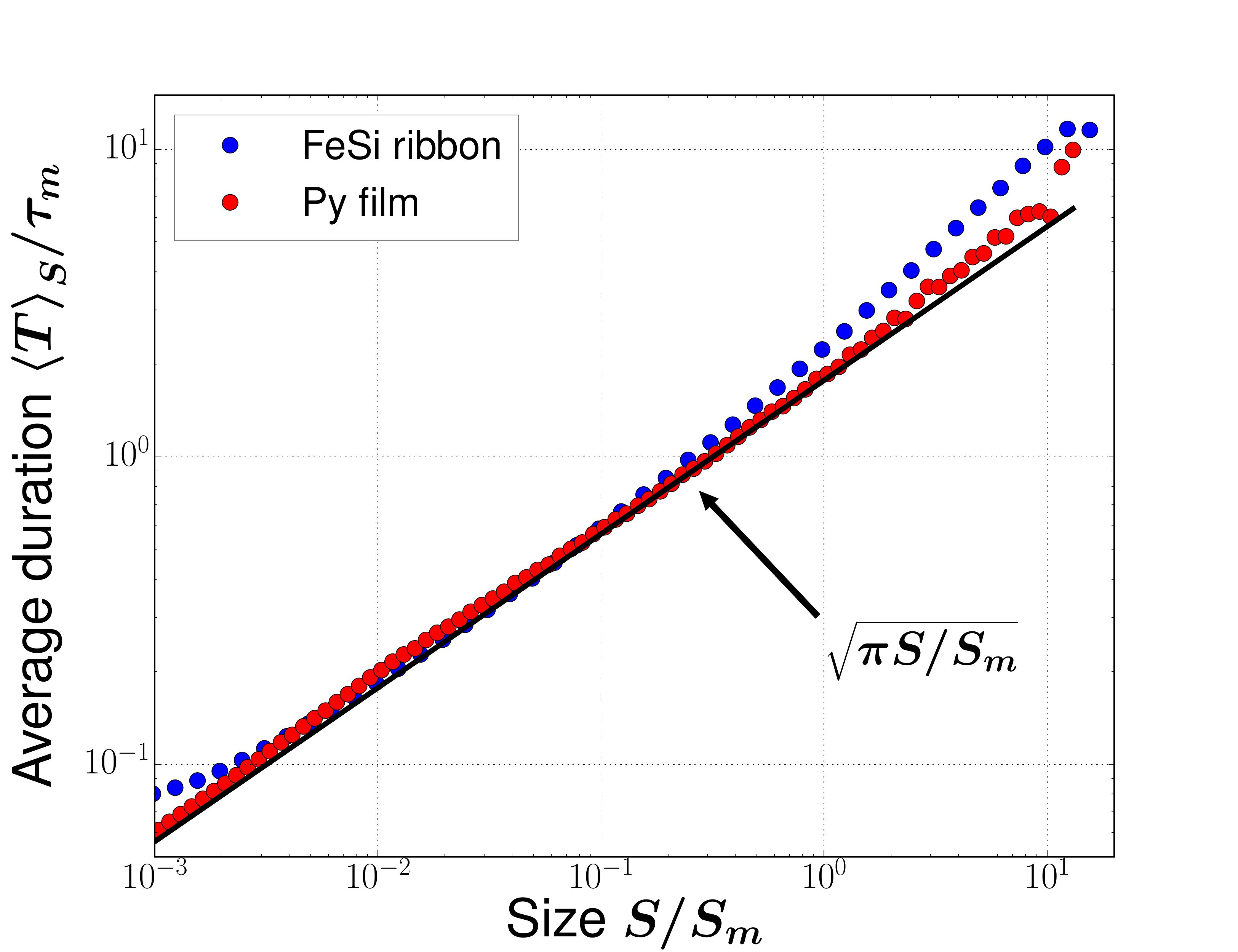}
\caption{Normalized average duration $\langle T \rangle_S/\tau_m $ of 
avalanches in the FeSi ribbon (blue dots) and the Py thin film (red dots) as a 
function of the normalized size $S/S_m$. 
The continuous line is the theoretical prediction of Eq.~(\ref{eq:TvsSMF}).}
\label{fig:aveT_MF}
\end{figure}

Let us now consider the average temporal avalanche shape at fixed size, $\langle \dot {\sf 
u}(t) \rangle_S$, which takes the form
\bea \label{eq:shapeS} 
\langle \dot {\sf u}(t) \rangle_S = \frac{S}{\tau_m} 
\Big(\frac{S}{S_m}\Big)^{\!-\frac{1}{\gamma}} f\bigg( 
\frac{t}{\tau_m}\Big(\frac{S_{m}}{S}\Big)^{\!\frac{1}{\gamma}} \bigg),
\eea 
where $f(t)$ is a universal scaling function, and $\int_0^{\infty} \rmd t\, 
f(t) = 1$. In mean field, $f(t)$ is independent of $S/S_m$ 
\cite{DobrinevskiLeDoussalWiese2013}
and reads
\bea \label{eq:f_t0}
f_{\rm MF}(t) = 2 t e^{- t^2}.
\eea 
Beyond MF, the function $f(t)$ has been obtained to ${\cal O}(\varepsilon)$ for 
SR elasticity. Here, we use the convenient form
\bea \label{eq:f_t}
f(t) \approx 2 t e^{-C t^\delta} B \exp\!\left(- \frac{\epsilon}{9} 
\!\left[\frac{\delta f(t)}{f_{\rm MF}(t)}{-}t^2 \ln (2t)\right]  \right),
\eea
where the function $\delta f(t)$ is displayed in Eq.~(34) of Ref.\
\cite{DobrinevskiLeDoussalWiese2014a}
and $B$ chosen s.t.\ $\int_0^{\infty} dt~f(t) = 1$, an approximation exact to 
${\cal O}(\varepsilon)$. Eq.~(\ref{eq:f_t}) has  asymptotic behaviors  
\begin{flalign} 
\label{eq:asympt}
f(t) &\simeq_{t \to 0} 2 A t^{\gamma-1}, \\
f(t) &\simeq_{t \to \infty}  2 A' t^{\beta} e^{- C t^\delta} 
\label{eq:asympt2},
\end{flalign}
with $A=1 + \frac{\epsilon}{9} (1-\gamma_{\rm 
E})$, $A' = 1+ \frac{\epsilon}{36}  (5 - 3 \gamma_{\rm E} - \ln 4)$, $\beta = 
1-{\epsilon}/{18}$, $C = 1+ \frac{\epsilon}{9} \ln 2$, and $\delta = 2 
+ {\epsilon}/{9}$.

\begin{figure}[tb]
\includegraphics[width=8.5cm]{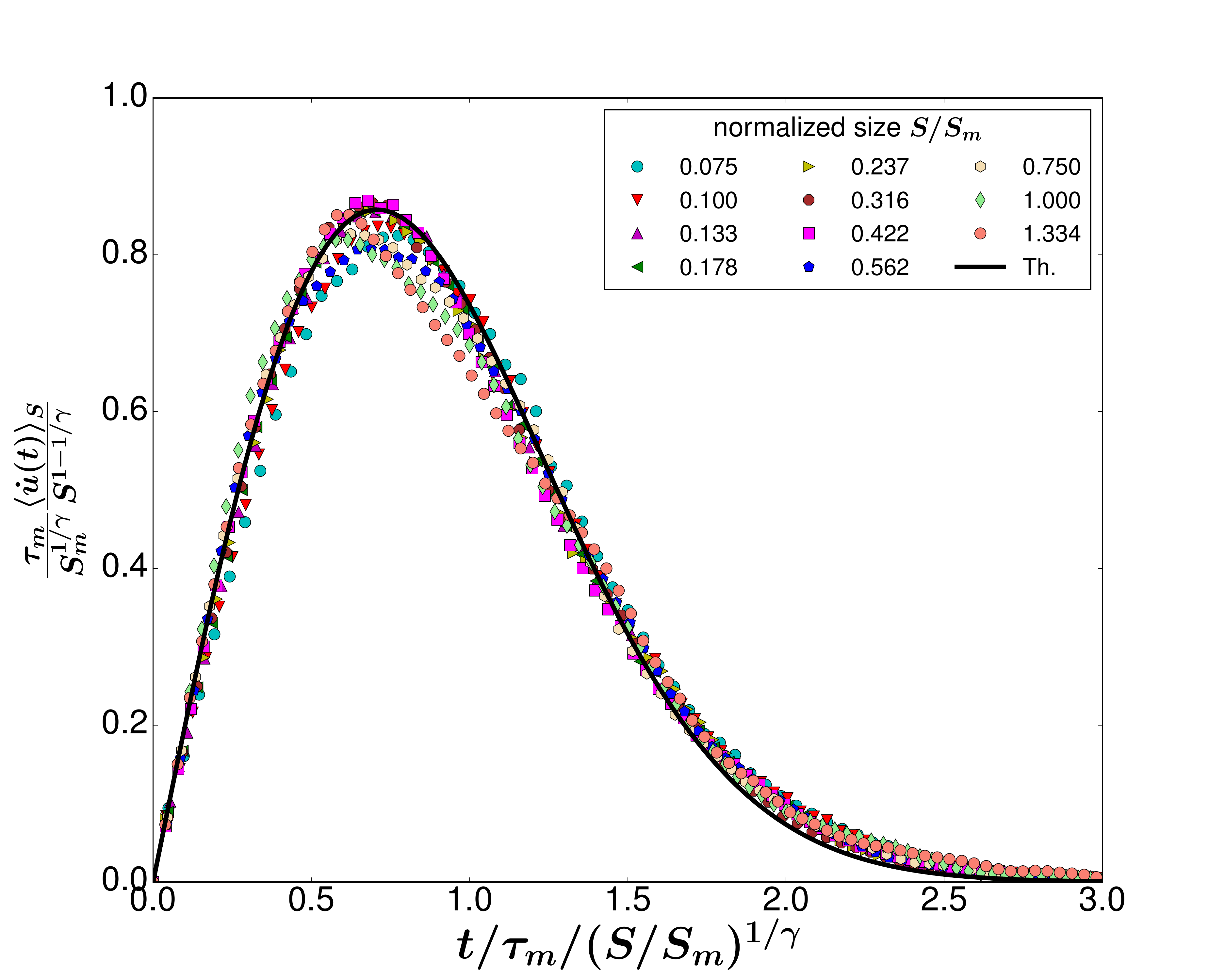}
\caption{Scaling collapse of the average shapes at fixed avalanche sizes 
$\langle \dot {\sf u}(t) \rangle_S$, according to Eq. (\ref{eq:shapeS}), in the Py 
thin film. The continuous line is the mean-field universal scaling function in 
Eq.~(\ref{eq:f_t0}).}
\label{fig:aveShape_Py}
\end{figure}

To compare these theoretical predictions to our experimental results, we first 
need to make use of dimensionless units, thus rescaling sizes and durations 
by $S_m$ and $\tau_m$, respectively. The parameter $S_m$ is analytically defined by 
Eq.~(\ref{eq:Sm}); we tested that it leads to a consistent
comparison of the measured size distribution $P(S)$ with the theory, see Supplemental Material~\cite{unp}. 
In particular, we verified that the cutoff occurs at $4 S_m$, as predicted. 
On the other hand, $\tau_m$ can only be inferred using (i) the 
expression of Eqs.~(\ref{eq:aveS}) and (\ref{eq:TvsSMF}); and
(ii) the data of the average shape $\langle \dot {\sf 
u}(t) \rangle_S$. 
In absence of ECs, the estimation of $\tau_m$ is made by matching 
both kind of data with the analytical expressions, giving a consistent and 
robust estimation of the parameter~\cite{tau_m}. In presence of ECs, we use the
procedure (i) to estimate $\tau_m$. 

\medskip

We analysed the avalanche statistics in different classes of materials. Two of 
them are thin films with negligible eddy current effects: a LR polycrystalline 
Ni$_{81}$Fe$_{19}$ Permalloy (Py) with a thickness of $200$ nm 
($\tau_m=39\, \mu$s) and a SR amorphous Fe$_{75}$Si$_{15}$B$_{10}$ (FeSiB) alloy, with a thickness of $1000$ nm 
($\tau_m=38\, \mu$s)~\cite{BOH-14, BOH-13}. The other two samples are ribbons with 
a thickness of about $20\,\mu$m, where eddy current retarding effects are well 
known: a LR polycristalline FeSi alloy with Si=$7.8$\%, ($\tau_m=2$ ms), and a SR amorphous 
Fe$_{64}$Co$_{21}$B$_{15}$ (FeCoB) alloy, measured under a small tensile stress of $2$ MPa ($\tau_m=0.5$ ms)~\cite{DurinZapperi2000, DUR-06}. 
All samples have a space dimension $D=2+1$; the two LR materials show 
MF exponents, while the other ones have NF exponents, with $\epsilon = 2$. Further details 
on the samples and the experiments are given in the 
Supplemental Material \cite{unp}.

Figures~\ref{fig:aveS_MF} and \ref{fig:aveS} report the average size as a function 
of avalanche duration for LR and SR samples, respectively, compared to the 
theoretical prediction of Eqs.~(\ref{eq:f1_MF}) and~(\ref{eq:gamma}). In the 
absence of eddy currents, the correspondence is almost perfect, except for the 
highest $(S,T)$ values. For LR samples (Fig.~\ref{fig:aveS_MF}), the mean-field
prediction (\ref{eq:f1_MF}) crosses over from $\sim \tilde T^2$ to $\sim \tilde T$ at large avalanche
sizes, a trend which seems to agree with our data. It is 
often argued that a linear dependence can also arise from the
superposition of a multiplicity of active DWs. Indeed 
some of the largest avalanches are a superposition of 
smaller avalanches occurring in different parts of the sample, triggered by the 
relatively large change of the magnetization \cite{WHI-03}. 
Furthermore, the retarding effect of eddy currents 
makes large avalanches (say, for $S > S_m$) even longer, so that 
the average size further deviates 
from the theoretical prediction, especially in samples with more EC,
as seen from Fig.~\ref{fig:aveS_MF}. 
Note that the agreement with the MF predictions is also quite good 
{\it at the level of fluctuations} ({\it i.e.} the second moment
$\langle S^2 \rangle_T$ in the inset of Fig.~\ref{fig:aveS_MF}). 
For the SR samples in Fig.~\ref{fig:aveS}, we plot the 
prediction for small $\tilde T$, in good agreement
with the data up to the size cutoff $4S_m$, {\it i.e.} $\langle S \rangle_T /S_m \sim 4$. At large $\tilde T$ we expect a similar bending to a linear 
behavior, although there are presently no detailed predictions for the crossover.

The mean avalanche duration at fixed size, $\langle T \rangle_S$, is shown in Fig.~\ref{fig:aveT_MF} for LR samples. For the film, it shows
an almost perfect agreement with the MF prediction of Eq.~(\ref{eq:TvsSMF}),
indicating that ECs are indeed negligible, while the effect of ECs is clearly 
visible in the ribbon. 

\begin{figure}[tb]
\includegraphics[width=8.5cm]{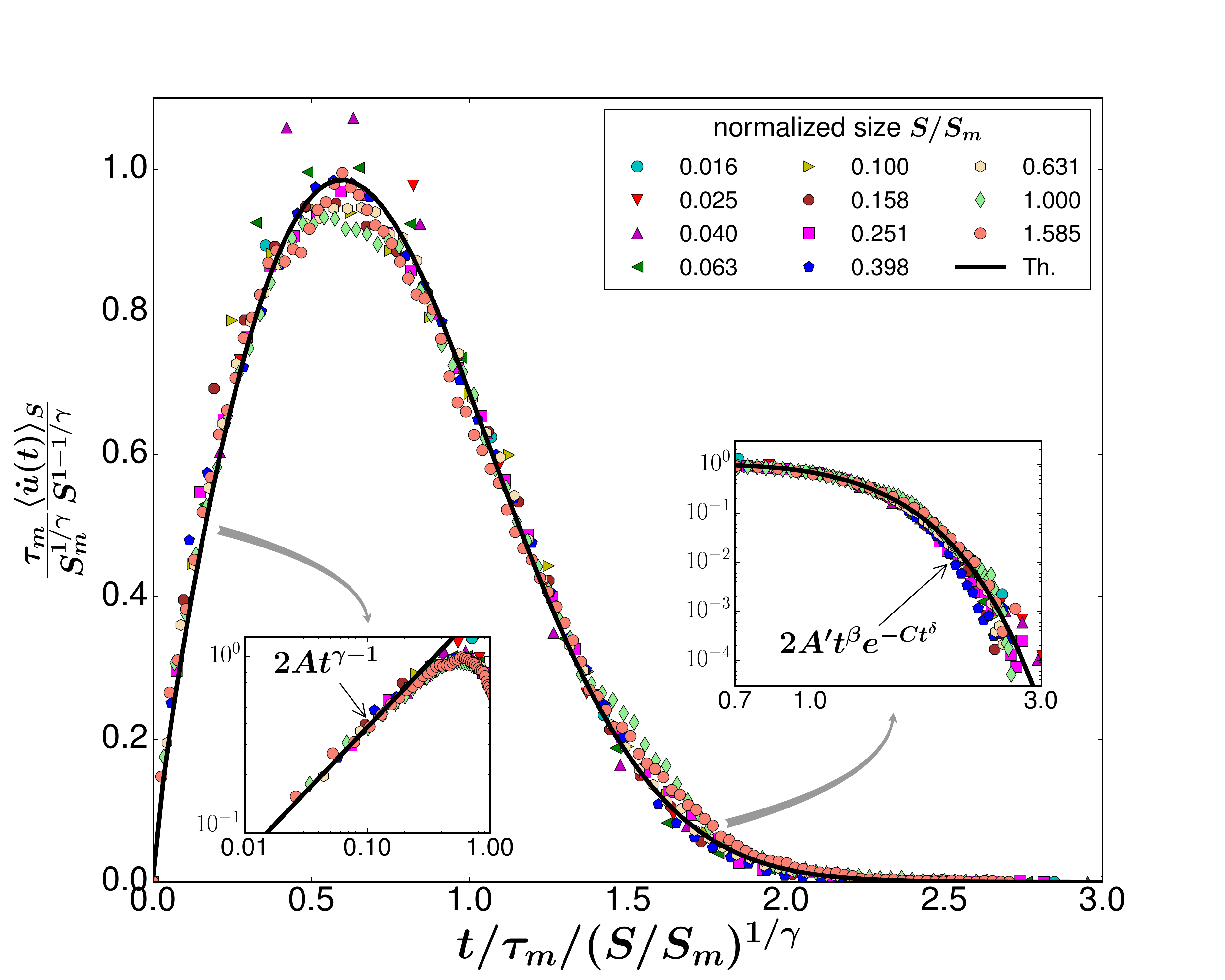}
\caption{Scaling collapse of the average shape at fixed avalanche sizes 
$\langle \dot {\sf u}(t) \rangle_S$, according to Eq.\ (\ref{eq:shapeS}), in the FeSiB thin film. 
The continuous line is the prediction for the universal SR scaling function of 
Eq.~(\ref{eq:f_t}). 
The insets show comparisons of the tails of the data
with the predicted asymptotic behaviors of 
Eqs.~(\ref{eq:asympt}) and (\ref{eq:asympt2}), setting $\epsilon=2$, with 
$A=1.094, A'=1.1, \beta=0.89, C=1.15$, and $\delta=2.22$.}
\label{fig:aveShape_FeSiB}
\end{figure}

Collapsing the experimental data of the average shapes at 
fixed size $\langle \dot {\sf u}(t) \rangle_S$ gives an 
alternative powerful way to estimate the exponent $\gamma$, as reported 
in Figs.~\ref{fig:aveShape_Py} and \ref{fig:aveShape_FeSiB}. Here we obtain the same 
exponents as from the average size measurements $\langle S \rangle_T$, {\it i.e.} 
$\gamma = 2$, and $\gamma = 1.76$, for LR and SR respectively. The collapsed 
average shapes correspond remarkably well to the theoretical predictions of 
Eqs.~(\ref{eq:f_t0}) and~(\ref{eq:f_t}), including the behavior in the tails 
(shown in the SR case in the insets of Fig.\ \ref{fig:aveShape_FeSiB}).
In the Supplemental Material \cite{unp}, we further verify that neither the collapse,
nor the quantitative fit can be achieved using the MF prediction.

\begin{figure}[tb]
\includegraphics[width=8.5cm]{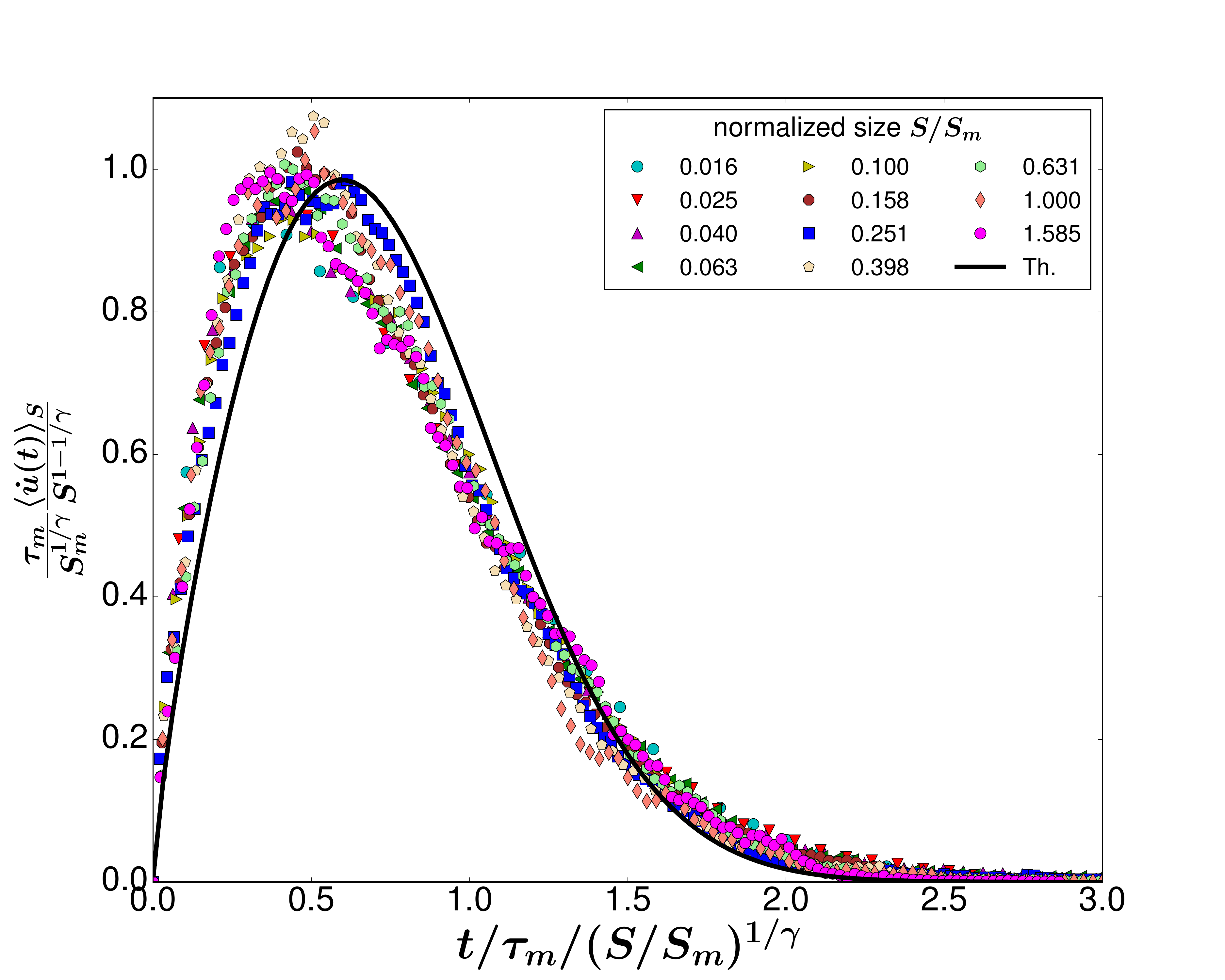}
\caption{Scaling collapse of the average shape at fixed avalanche sizes 
$\langle \dot {\sf u}(t) \rangle_S$ in the FeCoB ribbon using the  the 
theoretical values of $\gamma$ and $\tau_m$, as in Fig.~\ref{fig:aveS}.
The collapse deviates from the universal functions predicted for SR systems 
(continuous black line) in absence of eddy currents.}
\label{fig:aveShape_F64}
\end{figure}

Finally, it is well known that  relaxation of  eddy currents introduces a slow time 
scale into the dynamics, stretching avalanches  in time 
\cite{DobrinevskiLeDoussalWiese2013}. In Fig.~\ref{fig:aveShape_F64},
we have obtained an approximate collapse for the SR case in 
presence of eddy currents, using the theoretical value of $\gamma = 1.76$.
It is a manifest that the resulting curve is different from the one predicted in 
absence of  retardation effects.
Hence,  this is another unambiguous method to detect the presence of ECs, 
similarly to the leftward asymmetry of the temporal avalanche shapes at fixed 
durations \cite{ZAP-05}.
To go further and obtain predictions for the average shape in presence of ECs is 
difficult, as the shape strongly depends on the detailed parameters of the eddy 
currents. A step in that direction was obtained within MF in Ref.~\cite{DobrinevskiLeDoussalWiese2013} for a particular model of retardation. 
Detailed comparison with experiments involve non-universal scales, and is 
left for a future pubblication.

In conclusion we have shown how the data from Barkhausen noise experiments can 
be analyzed and confronted to the most precise recent 
theoretical predictions. This provides very quantitative and fundamental tests of the
theory of avalanches beyond scaling exponents. The prediction of
universality will also lead to a better characterization of magnetic systems,
allowing to measure its dimensionality and, via the deviations from the 
theory, effects of multiple avalanches or eddy currents.

\acknowledgments
This work was supported by PSL grant ANR-10-IDEX-0001-02-PSL and CNPq (Grants 
No.~$306423$/$2014$-$6$, No.~$471302$/$2013$-$9$, No.~$306362$/$2014$-$7$, and 
No.~$441760$/$2014$-$7$). 
We acknowledge hospitality of the KITP with 
support in part by the National Science Foundation 
under Grant No. NSF PHY11-25915.


\begin{thebibliography}{34}
\bibitem{tau_m}
Its actual value depends on the
threshold which defines the start of the avalanche \cite{DUR-06},
however the latter can safely be set within a range where the critical 
exponents' estimation is robust, as explained in~\cite{LAU-09a}.
\end{thebibliography}



\def\figwidth{8cm}

\renewcommand{\thefigure}{S\arabic{figure}}

\setcounter{figure}{0}

\newpage

\begin{figure*}[bth]
\includegraphics[width=8.5cm]{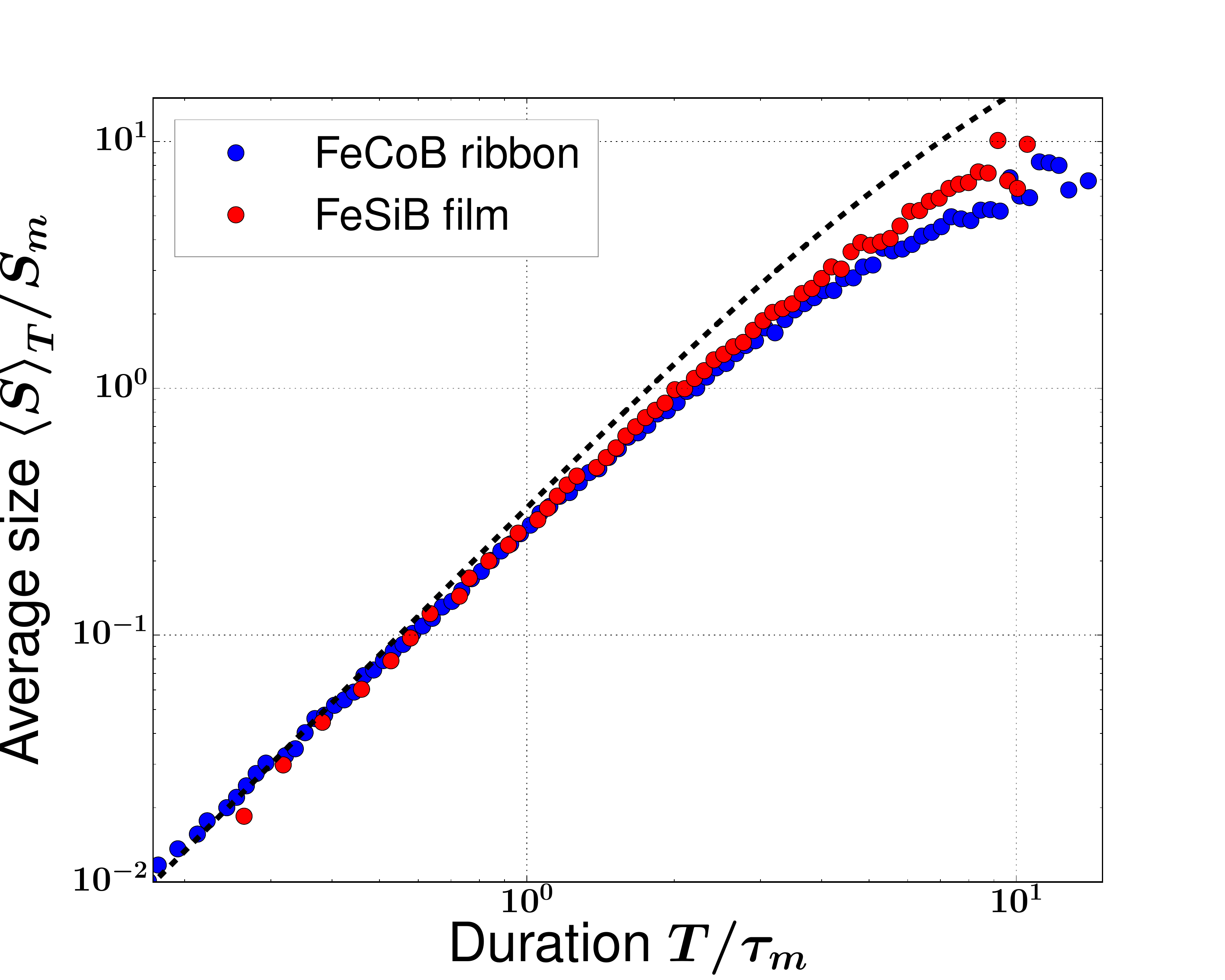}\includegraphics[width=8.5cm]{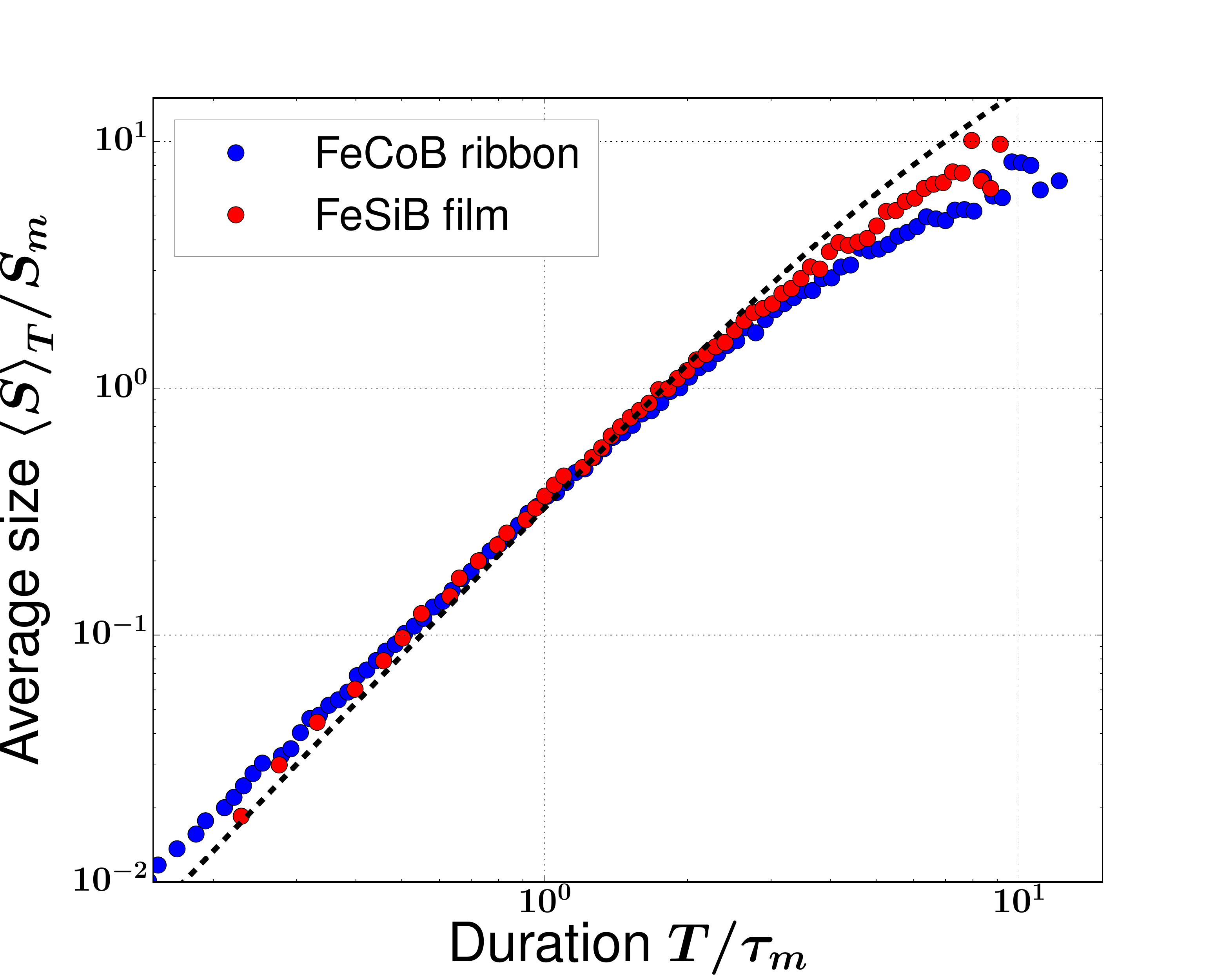}
\caption{Average sizes $\langle S \rangle_T/S_m$ of avalanches in the SR 
samples, FeCoB ribbon (blue dots) and the FeSiB thin film (red dots). The dash 
lines are the theoretical MF prediction $g_1^{\rm MF}(\tilde T)=2 \tilde T 
\coth(\tilde T/2)-4$, for two different values of $\tau_m$.}
\label{fig:fig2MF}
\end{figure*}

\begin{figure*}[tbh]
\includegraphics[width=8.5cm]{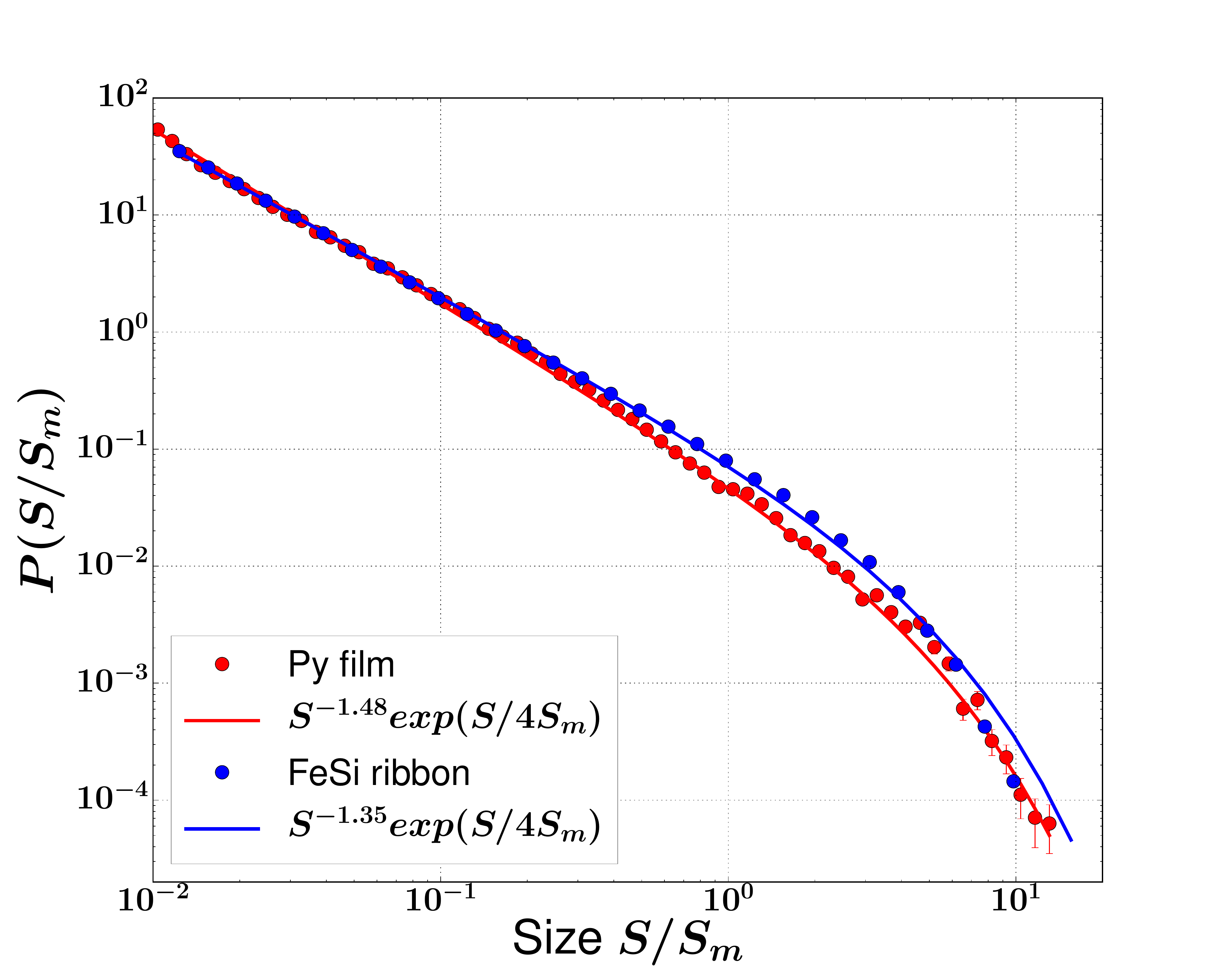}\includegraphics[width=8.5cm]{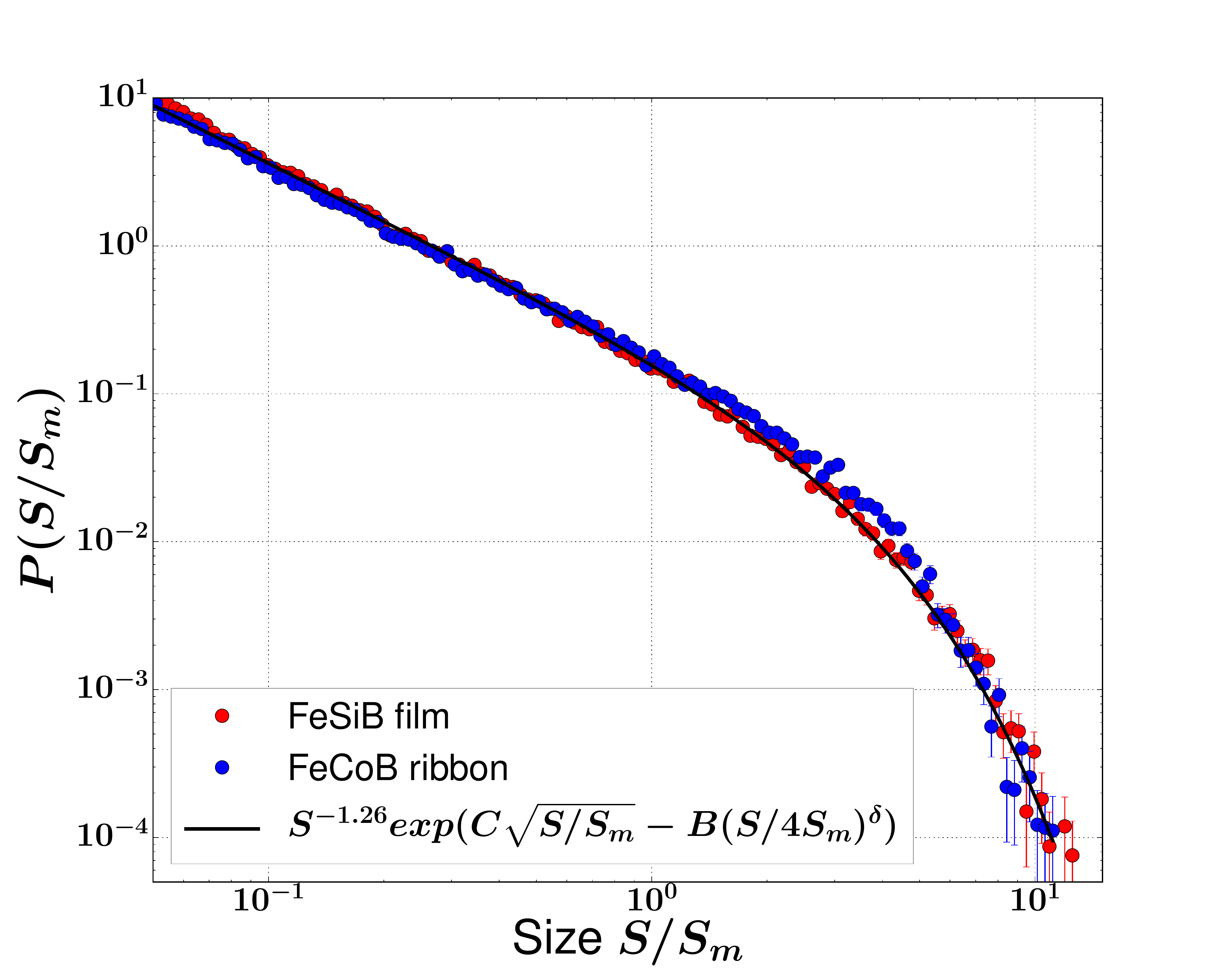}
\caption{Left: Distributions of avalanche sizes $P(S)$ in MF samples, the FeSi 
ribbon (blue dots) and the Py thin film (red dots). 
The continuous line is the theoretical prediction $P(S) = S^{-\tau} 
\exp(S/4S_m)$ of Ref.~\cite{RossoLeDoussalWiese2009a}. Here, $\tau$ is a little 
smaller of the adiabatic value of $1.5$ because of the finite rate 
effects~\cite{PapanikolaouBohnSommerDurinZapperiSethna2011}. On the other hand, 
the cutoff of the distribution is perfectly predicted at $4S_m$. Right: $P(S)$ 
of SR samples, the FeCoB ribbon (blue dots) and the FeSiB thin film (red dots). 
The continuous line is the theoretical prediction of 
Ref.~\cite{LeDoussalWiese2008c}, Eqs.\ ($169$)-($172$), in which 
$P(S) = S^{-\tau} \exp(C\sqrt{S/S_m} -B (S/4S_m)^\delta)$, with $\tau=1.26$, $B 
= 1.51$, $C = 0.34$, and $\delta = 1.11$.	}
\label{fig:PS}
\end{figure*}

\begin{figure*}[tbh]
\includegraphics[width=8.5cm]{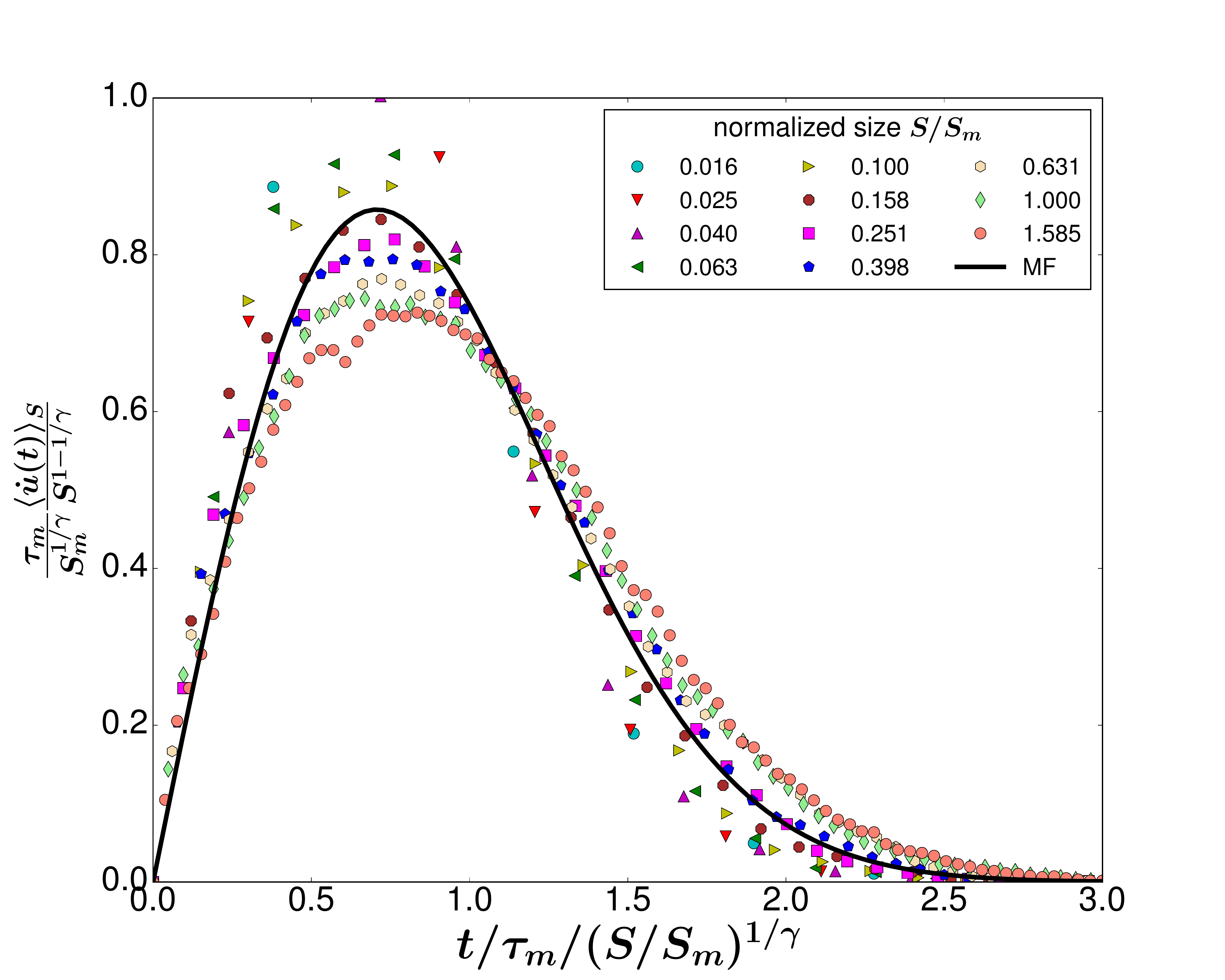}\includegraphics[width=8.5cm]{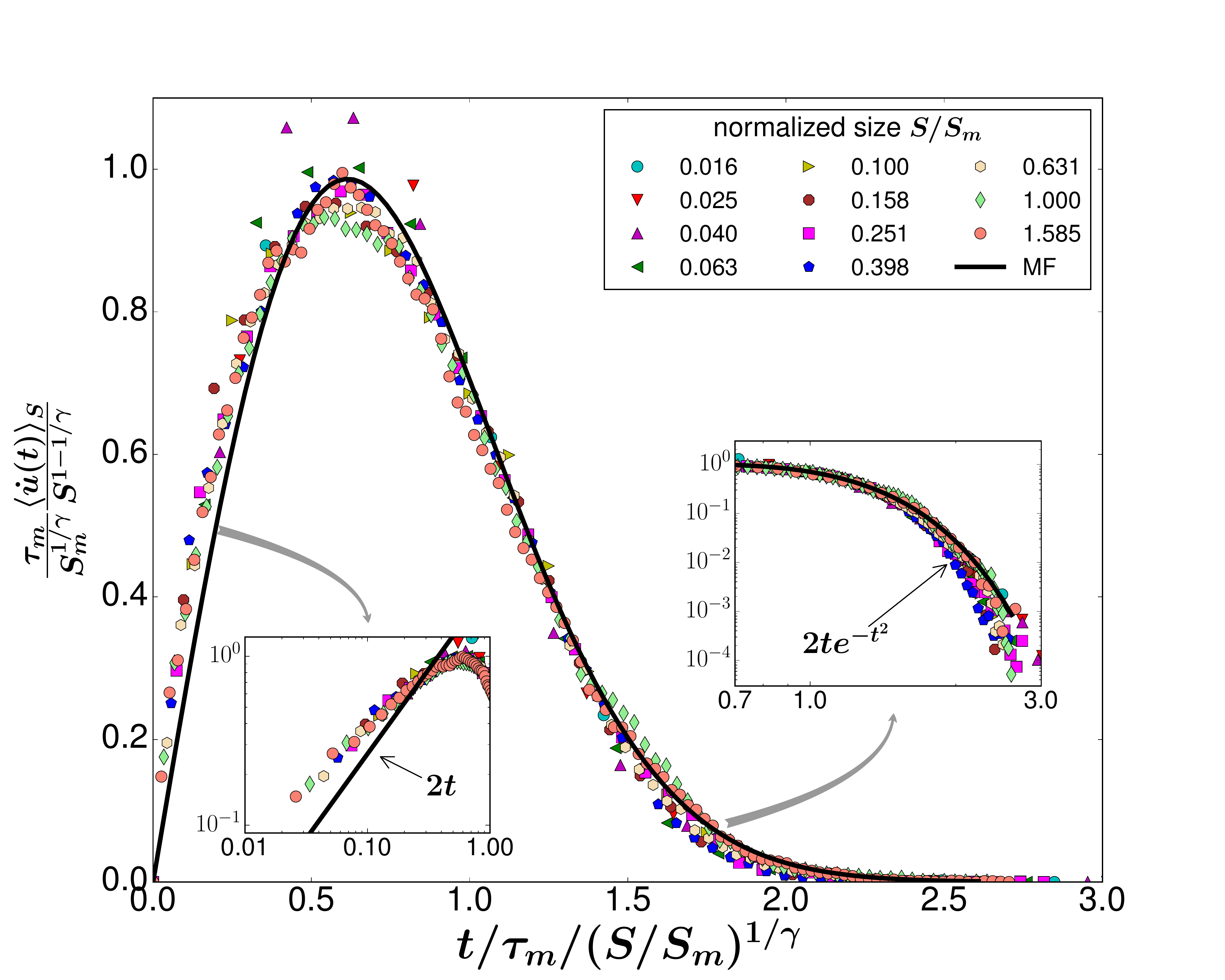}
\caption{Average shapes for the SR samples, rescaled with the MF exponent 
$\gamma = 2$ (left), and with the theoretical SR value $\gamma = 1.76$ compared 
to the MF scaling function $f_{MF}(t) = 2 t e^{-t^2}$ (right). While in the 
former the data do not collapse at all, as expected, in the latter the MF 
prediction can roughtly describe the experimental data once a $\tau_m$ is 
arbitrary set. Here we use a \% 15 higher than the SR value. As a matter of 
fact, the MF prediction does not describe the data at low values, as the 
function is strictly linear, while the data follow a $t^{\gamma-1}$ law.}
\label{fig:fig5MF}
\end{figure*}

\begin{center}
{\bf Supplemental material of \\Quantitative scaling of magnetic avalanches}

G.~Durin, F.~Bohn, M.~A.~Corr\^{e}a, R.~L. Sommer, P.~Le Doussal, K.~J.~Wiese
\end{center}

\vspace{.5cm}
In this supplemental material, we 
\begin{enumerate}
\item[(A)] Provide information on the investigated samples and details on the 
experiments;
\item[(B)] Further characterize the samples, and discuss distributions of 
avalanche sizes and universality classes;
\item[(C)] Provide additional comparison of the data with the theoretical 
mean-field predictions.
\end{enumerate}

\vspace{.5cm}
{\bf A. Samples and experiments}
\vspace{.25cm}

We employ thin films and ribbons to perform Barkhausen noise experiments. 

The thin films correspond to a polycrystalline Ni$_{81}$Fe$_{19}$ Permalloy (Py) 
film with thickness of $200$~nm and an amorphous Fe$_{75}$Si$_{15}$B$_{10}$ 
(FeSiB) film with the thicknesses of $1000$~nm. The films are deposited by 
magnetron sputtering onto glass substrates, with dimensions $10$ mm $\times$ $4$ 
mm, covered with a $2$ nm thick Ta buffer layer. The deposition process is 
carried out with the following parameters: base vacuum of $10^{-7}$ Torr, 
deposition pressure of $5.2$ mTorr with a $99.99$\% pure Ar at $20$ sccm 
constant flow. The Ta layer is deposited using a DC source with current of $50$ 
mA, while the ferromagnetic layers are deposited using a $65$ W RF power supply. 
During the deposition, the substrate moves at constant speed through the plasma 
to improve the film uniformity, and a constant magnetic field of $1$ kOe is 
applied along the main axis of the substrate in order to induce magnetic 
anisotropy. Detailed information on the structural and magnetic 
characterizations is found in Refs.~\cite{BOH-14,BOH-13}. 

The ribbons are a polycrystalline FeSi alloy with Si=7.8\% and an amorphous 
Fe$_{64}$Co$_{21}$B$_{15}$ (FeCoB), both with thickness of about $20\,\mu$m. The 
experiments are performed with ribbons of dimensions of about $20$~cm $\times 
1$~cm, and the FeCoB one is measured under a small tensile stress of $2$ MPa in 
order to enhance the signal-noise ratio. Further information on the ribbons is 
found in Refs.~\cite{DurinZapperi2000,DUR-06}.

We record Barkhausen noise time series using the traditional inductive technique 
in an open magnetic circuit, in which one detects time series of voltage pulses 
with a sensing coil wound around a ferromagnetic material submitted to a 
slow-varying magnetic field. In our setup, the sample and pick up coils are 
inserted in a long solenoid with compensation for the borders, to ensure an 
homogeneous applied magnetic field on the sample. The sample is driven by a 
triangular magnetic field, applied along the main axis of the sample, with an 
amplitude high enough to saturate it magnetically. Barkhausen noise is detected 
by a sensing coil wound around the central part of the sample. A second pickup 
coil, with the same cross section and number of turns, is adapted in order to 
compensate the signal induced by the magnetizing field. The Barkhausen signal is 
then amplified, filtered, and finally digitalized. For the thin films, 
Barkhausen noise measurements are performed using a sensing coil with $400$ 
turns, $3.5$ mm long and $4.5$ mm wide, and under similar conditions, {\it i.\ 
e.}, $50$~mHz triangular magnetic field, $100$ kHz low-pass filter set in the 
preamplifier and signal acquisition with sampling rate of $4\times 10^6$ samples 
per second~\cite{BOH-14, BOH-13}. For the ribbons, the experiments are carried 
out using a sensing coil with $50$ turns, $1$~mm long and $1$~cm wide, 
triangular magnetic field with frequency of $3$-$5$~mHz, and low pass 
preamplifier filter chosen in the $3$-$20$~kHz range, roughly half of the 
sampling rate~\cite{DurinZapperi2000, DUR-06}. All the time series are acquired 
just around the central part of the hysteresis loop, near the coercive field, 
where the domain wall motion is the main magnetization mechanism~\cite{DUR-06}. 
For each experimental run, the statistical properties are obtained from at least 
$150$ measured Barkhausen noise time series.

\vspace{.5cm}
{\bf B. Distribution of avalanche sizes} 
\vspace{.25cm}

Here, we further characterize the investigated samples and show why they belong 
to different universality classes. Figure~\ref{fig:PS} presents the 
distributions of avalanche sizes $P(S)$ for the LR and the SR samples. All the 
distributions follow a cutoff-limited power-law scaling behavior, as expected. 
The LR samples, {\it i.e.} the polycrystalline Py thin film and FeSi ribbon, are 
characterized by mean-field exponent, $\tau \sim 1.5$, and well described by the 
theoretical prediction $P(S) = S^{-\tau} \exp(-S/4S_m)$ of 
Ref.~\cite{RossoLeDoussalWiese2009a}. Due to finite rate effects, the 
experimental exponents are a little smaller than the prediction, although the 
cutoff of the distribution is perfectly predicted at $4S_m$. 
The SR samples, {\it i.e.} the FeSiB thin film and the FeCoB ribbon, present the 
Narayan-Fisher exponent $\tau=1.26$, and are well described by the theoretical 
prediction of Ref.~\cite{LeDoussalWiese2008c}, Eqs.\ ($169$)-($172$), in which
$P(S) = S^{-\tau} \exp(C\sqrt{S/S_m} -B (S/4S_m)^\delta)$, with $B = 1 + 
\alpha(1+\gamma_E/4)$, 
$C = 1/2\sqrt{\pi}\alpha$, $\alpha = 1/3(1-\zeta_1)\epsilon$, $\delta = 1+ 
\epsilon/18$, where $\zeta_1 = 1/3$ and $\epsilon = d_{uc} -d = 2$.

\vspace{.5cm}
{\bf C. Additional comparison with mean-field predictions} 
\vspace{.25cm}

The aim of this section is to show that the data for the SR samples cannot be 
fit consistently with the mean-field predictions.

Let us start with the average normalized size $\langle S \rangle_T/S_m$ as a 
function of the normalized duration $\tilde T = T/\tau_m$. When compared with 
the MF prediction $g_1^{\rm MF}(\tilde T)=2 \tilde T \coth(\tilde T/2)-4$, we 
need to adapt the value of the $\tau_m$, as explained in the main paper. Here we 
need to increase its value by 30\% (see Fig.~\ref{fig:fig2MF}, left) up to 50\% 
(right) compared to the value of Fig.~2 of the paper. It is clear that the MF curve, having 
$\gamma = 2$, larger than the experimental data, either follows low value data 
but fails for the higher ones or intersects them at some point. In practice, it 
cannot correctly describe the experiments.

Let us now discuss the data for the avalanche shapes. 

The main question is whether the data for the SR samples can be fitted by the MF 
predictions, which means (i) $\gamma=2$, and (ii) a scaling
function $f(t) = f_{MF}(t) = 2 t e^{-t^2}$. In Fig.~\ref{fig:fig5MF} (left) we 
present the data rescaled 
with the choice $\gamma=2$. Obviously there is no data collapse for various 
sizes. This feature is clearly independent of the choice of $\tau_m$ 
as depends on $\gamma$ only. This rules out the MF prediction to compare with the 
data. 

Another question, mostly of theoretical interest, is how the scaling function
of the rescaling of the data with $\gamma \sim 1.76$ in Fig. 5 of the main paper 
differs quantitatively from the MF prediction
$f_{MF}(t)$. In Fig.~\ref{fig:fig5MF} (right) to compare the two curves we need 
to artificially change the $\tau_m$ value, which is smaller and incompatible 
with the values of Fig.~\ref{fig:fig2MF}. More striking, the scaling function is 
linear at low values, while the data follow a $t^{\gamma-1}$ law.

In conclusion, it is important to emphasize that a quantitative comparison of the experimental data 
with the theoretical predictions is only possible when the values of the parameters $\gamma, \tau_m, S_m$ are fixed, and compatible with the theoretical curves. 
In other words, a single visual comparison may not be enough to establish the universality class. Here we demonstrated that the use of at least two independent measurements (such as the average size, 
and the average shapes at fixed sizes) set constrains that unambiguously determine the universality class. This is particularly important when applied to other systems showing a similar crackling noise, 
or in more general displaying a dynamical phase transition.


\vspace*{0.9cm}

\end{document}